\documentclass[preprint]{aastex}
\usepackage{psfig}

\newcommand{\hal}{H$\alpha$}
\newcommand{\mj}{M$_{Jup}$}
\newcommand{\kms}{km s$^{-1}$}
\newcommand{\delv}{$\Delta$v}

\slugcomment{Submitted to AJ}

\shortauthors{Basri \& Mart\'\i n}
\shorttitle{First Brown Dwarf Binary}

\begin{document}

\title{PPl~15: The First Brown Dwarf Spectroscopic Binary}

\author{Gibor Basri} 
\and 
\author{Eduardo L. Mart\'\i n}
\affil{Astronomy Department, University of California,
    Berkeley, CA 94720}

\email{basri@soleil.berkeley.edu, ege@popsicle.berkeley.edu} 

\begin{abstract} 

PPl~15 is the first object to have been confirmed as a brown dwarf by the 
lithium test (in 1995), though its inferred mass was very close to the 
substellar limit. It is a member of the Pleiades open cluster. Its position in a 
cluster color-magnitude diagram suggested that it might be binary, and 
preliminary indications that it is a double-lined spectroscopic binary were 
reported by us in 1997. Here we report on the results of a consecutive week of 
Keck HIRES observations of this system, which yield its orbit. It has a period 
of about 5.8 days, and an eccentricity of 0.4$\pm0.05$. The rotation of the 
stars is slow for this class of objects. Because the system luminosity is 
divided between 2 objects with a mass ratio of 0.85, this renders each of them 
an incontrovertible brown dwarf, with masses between 60-70\mj. We show that 
component B is a little redder than A by studying their wavelength-dependent 
line ratios, and that this variation is compatible with the mass ratio. We 
confirm that the system has lithium, but cannot support the original conclusion 
that it is depleted (which would be surprising, given the new masses). This is a 
system of very close objects which, if they had combined, would have produced a 
low mass star. We discuss the implications of this discovery for the theories of 
binary formation and formation of very low mass objects.

\end{abstract}

\keywords{open clusters and associations: individual (Pleiades) 
--- stars: low-mass, brown dwarfs 
--- stars: binaries  
--- stars: formation}

\newpage
\section{Introduction}

Basri, Marcy \& Graham (1996; hereafter \citeauthor{bmg}) obtained in 1995 the first unambiguous detection of lithium in a brown dwarf (BD) candidate.  This
object was PPl~15, discovered in a CCD survey by \citet{stauffer94}. 
In spite of the detection of lithium, the position of PPl~15 in the
color-magnitude diagram was very close to the expected substellar limit
for the newly proposed age of the cluster ($\sim$115~Myr).  It was possible
that PPl~15 could be a very low-mass (VLM) young star or a transitional object
rather than a genuine brown dwarf.  \citeauthor{bmg} inferred a mass of 
0.078~M$_\odot$, which was slightly below the substellar limit at the time 
(0.080~M$_\odot$), but is slightly above the latest expected substellar limit 
for solar metallicity \citet[0.075~M$_\odot$~ or 75\mj]{baraffe98}.

It has long been known that there is a dispersion in the Pleiades color-magnitude sequence, greater than the observational errors.  For some time this was thought to be due to an age spread in the cluster, but strong arguments have been brought against that \citep{stauffer95,steele95}.  It is now thought that the dispersion is likely due to unresolved binaries.  These are not only expected, but explain the amplitude of the spread quite naturally.  PPL 15 sits at a location which suggests that it is one of these binaries \citep{osorio97}.

An HST/WFPC2 image (J. McConnell \& J. Surdej; private communication) eliminates 
companions to within 0.15" (18 AU) of the primary (with a contrast up to 5 
magnitudes at I). Spectroscopically, \citeauthor{bmg} reported only that the mean line positions did not change at the level of 5 \kms~ (1 pixel),
then coadded the spectra to analyze lithium.  We re-examined \citeauthor{bmg}'s
HIRES data to look for radial velocity variations using a full
cross-correlation analysis of each night.  To our great surprise, the
cross-correlation peaks are double with comparable amplitudes, except on the
first night (where there is a single peak).  We thus reported that PPl~15 is a 
likely double-lined spectroscopic binary \citet{basmar98}.  In this paper
we present a set of new HIRES observations which confirm the binarity of PPl~15
and allow us to derive its orbital parameters.  We also find an \hal~ flare on 
one of the nights, and confirm the lithium detection reported by \citeauthor{bmg}.  We discuss the implications of our results for the binary frequency among BDs and the theories of binary formation.

\section{Observations}

Our full set of observations of PPl~15 were obtained at 3 different epochs, all 
at the Keck Observatory. The first 6 observations on 23-25 Nov. 1994, were the 
initial set in \citeauthor{bmg} which was aimed at conducting the lithium test. These 
were consecutive pairs of exposures on 3 consecutive nights. On 5-6 March 1995 
\citeauthor{bmg} observed it four more times (again in consecutive pairs) attempting 
to confirm the initial preliminary detection of lithium. After analysis of these 
observations for binarity in early 1997 \citep{basmar98}, we conducted 9 
observations on 1-7 Dec. 1997 to determine the orbit. In this latter run we are 
grateful for the cooperation of Geoff Marcy and Paul Butler, and John Stauffer 
and Robert O'Dell, who allowed us to trade observations on their nights for time 
we had allocated. It was only through this cooperation that we were able to 
obtain the unprecedented consecutive coverage on the Keck I telescope. The 
weather was not good for several nights of the run, so several spectra were 
obtained through clouds and barely of useable quality. Fortunately, we only need 
sufficient signal for a cross-correlation analysis, and can use the reddest 
orders. The observing log for all observations is given in Table~1.

The observations were obtained at the Keck~I telescope, using the HIRES
echelle spectrometer \citep{vogt94}.  The instrumental configuration and data reduction procedure were the same as in \citeauthor{bmg}.  With a dispersion of about 0.01 nm per (binned) pixel, the resolving power obtained was R$\sim$31000. Standard echelle reductions including sky subtraction were done with our IDL routines.  Fully reduced data is available about 10 minutes after the spectrum is read out, enabling decisions about whether a further exposure is needed.  In addition to the spectra of PPl~15, a spectrum of Gl406 was taken with the same instrumental configuration to serve as a cross-correlation template.  The star HD 93521 (O9Vp) was also observed as a relative flux standard and to measure telluric absorption line contamination.

\section{Spectroscopic analysis}

Each order in each PPl~15 spectrum was first median filtered to remove cosmic 
rays and bad columns, then a continuum fit to our flux standard was divided out 
and the mean spectrum normalized to unity. At this point we used our IDL routine 
XCORL to obtain a cross-correlation function (XCF) between that order and the 
corresponding one in our velocity standard. The velocity standard for Nov. 1994 
and Mar. 1995 was LHS523, and for Dec. 1997 it was Gl406. A shift of up to 50 
pixels in either direction produced a nice XCF with one or two sharp peaks and 
plenty of flat ``continuum'' on either side of them. Because the S/N of the 
spectra was generally low and the orders contain varying amounts of spectral 
features which can produce a correlation signal, we averaged together XCFs of 
the 6 redmost orders and 2 others which produced the best XCFs. We avoided 
orders with a strong telluric signal. This final average XCF constitutes the 
``observation'' for each spectrum. We measured the positions of its peaks with a 
two-gaussian plus quadratic background fitting routine to obtain the velocity of 
each component in each spectrum. The errors in the individual velocity 
measurements are generally less than 3 \kms~ (though they vary with the quality 
of and shape of the XCF).

Because we used the same template spectrum for each observing epoch, we obtain 
good relative measures of the velocity shifts. The night sky airglow spectrum is 
also obtained in each observation, and we ensured that the template and program 
observation are on the same velocity scale using those. We then refer each 
velocity template to an absolute velocity reference by measuring the position of 
atomic lines in it compared with the night sky, and correcting for the solar 
system barycentric velocity. The systemic velocity of PPl~15 was consistent to 
within 2 \kms~ for each data set, and the overall system velocity was consistent 
with that published by \citeauthor{bmg} (though here we obtained a slightly higher 
radial velocity of 6.5$\pm 2$ \kms~ compared with their value of 4.6 \kms).

We were also able to obtain the projected rotation velocity for each star, by 
measuring the width of the XCF of each when they were widely separated. The 
method for deriving $v$ sin$i$~ from these was the same as in \citet{basri95}. 
The two stars appear to have the same $v$ sin$i$, each about 10 \kms. This is 
actually the lowest value (even after correction for inclination) we have found 
in Pleiades VLM stars \citep[see][]{oppheim97}, where velocities from 25-50 \kms~ 
are more common. One possibility is that the stars have been tidally 
pseudo-synchronized, meaning that the angular velocity of rotation is similar to 
the angular velocity of the companion crossing the sky during periastron 
passage. The timescale for this synchronization is much shorter than for orbital 
circularization (which we show below has not occurred). 

\begin{figure}{\label{lindep}}
\centerline{\psfig{figure=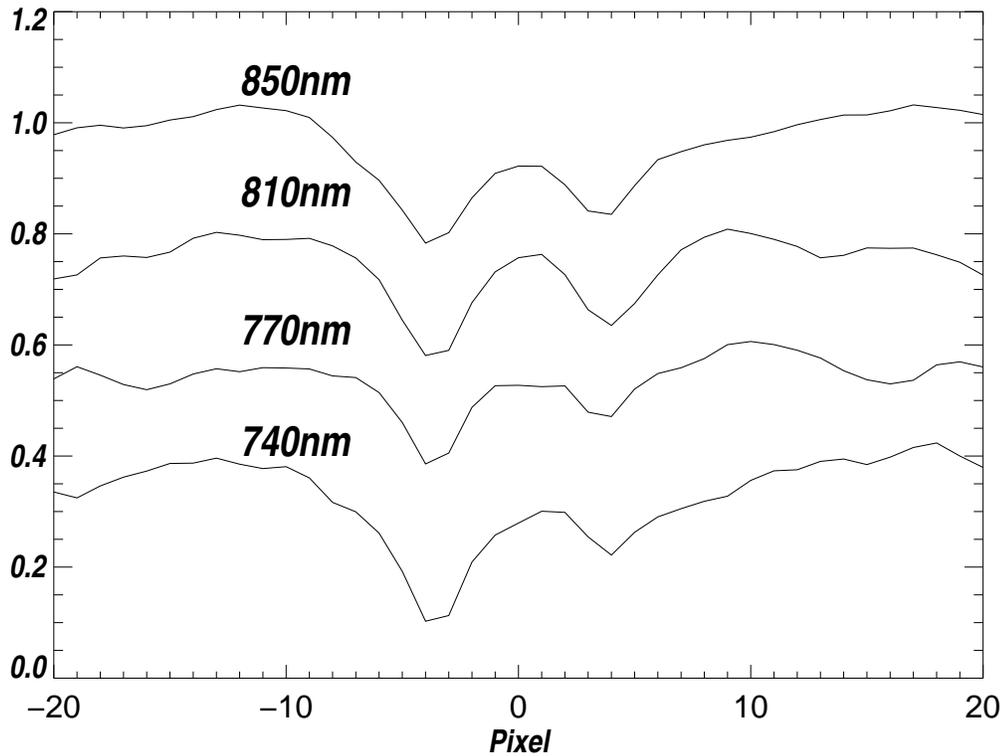,width=6in}}
\caption {The ratio of the average line depths as a function of wavelength, as found from a cross-correlation analysis. Various echelle orders from the spectra of 7 Dec 1997 are correlated with the same orders from Gl406 (our routine XCORL produces dips rather than peaks, which look more like the average spectral lines). Both components of PPl 15 are easily visible, as is the fact that the line ratio becomes more equal at longer wavelengths. } 
\end{figure}

\subsection{The masses of the components}

It is important to point out that we cannot derive the actual masses from the 
radial velocities we have measured (as we do not know the orbital inclination). 
We infer the total mass of the system (0.13M$_\odot$) using the fact that the 
components are almost equal in brightness (and therefore mass), then asking what 
mass they would have to be based on theoretical models such as those by 
\citet{baraffe98} to match the observed luminosity and color of the system. It 
was a similar analysis that led \citeauthor{bmg} to the conclusion that PPl~15 was at 
the substellar boundary in the first place. The total system mass is now almost 
doubled, because each component only drops about a factor of two in luminosity 
and the colors remain almost the same. 

There is no hope of spatially resolving the two components in the near 
future. Their mass ratio can be found from analysis of the individual 
velocities. For any arbitrary pair of phases, the mass ratio will be the ratio 
of the velocity difference for the primary divided by the velocity difference 
for the secondary. These are easily distinguished, because the 2 sets of lines 
are not of equal depth. We chose several pairs of phases in which the components 
had switched places, and measured the line positions by correlation with our 
velocity standards (after barycentric correction of all observations). Because 
of the division operation, the velocity ratio is quite sensitive to the measured 
centroids of the components (for example, a barycentric correction of only 1 
\kms~ for closely spaced observations changes that mass ratio determination by 
0.15). For the phase pairs RJD 50787.9/50789.9, 49681.9/49788.8, and 
49680.8/49789.9, we obtained mass ratios of 0.85, 0.91, and 0.82. We therefore 
set the mass ratio at 0.87$\pm0.04$ from this exercise.

We can also get some information on the mass ratio from the fact that the 
component lines, when separated, have a ratio of line depths that is wavelength 
dependent. Indeed the ratio becomes more similar at longer wavelengths, as 
expected if the fainter component is also redder. Given our 2-gaussian fits to 
the XCFs in each order, we can measure the ratio of the line depths. This was 
done for the average of the 7 Dec 1997 spectra, and the results are given in 
Fig. \ref{lindep}. The ratio of the brighter to fainter component goes from a 
little over 3 at 740 nm to 1.5 at 870 nm. The line depth ratios should 
approximately correspond to the continuum brightness ratios. The average 
spectral line depths themselves should not be a function of wavelength, 
especially given the small expected temperature difference between the 
components. 

To calibrate these results, we divided low dispersion spectra of Pleiades 
objects with spectral types M6-8 by each other. The ratios of the high 
resolution spectral lines between 770-790 nm and 810-850 nm are roughly 
consistent with the behavior of the divided low dispersion spectral shapes, as 
is the fact that the ratio is smaller for the redder wavelengths. The ratio at 
750 nm is perhaps greater than expected from the low dispersion spectra. 
Nonetheless, the general behavior is reasonably reproduced if the brighter 
component has a spectral class of M6 and the fainter is M7. This is consistent 
with the composite spectral type of M6.5 that has been given for PPl~15 
\citep{mart96}. The mass ratio that would be expected between those spectral 
types in the Pleiades is 0.8-0.9. Finally, the I magnitude difference between 
objects with 70\mj~ and 60\mj~ is predicted by \citet{baraffe98} to be 0.6 
magnitudes or a factor of 1.75 in brightness, which is also about the brightness 
ratio found from the line depths. We therefore adopt a mass ratio of 0.85 and 
individual masses (based on luminosity) of 70\mj~ and 60\mj~ for PPl~15 A \& B.

\begin{figure}{\label{orbit1}}
\centerline{\psfig{figure=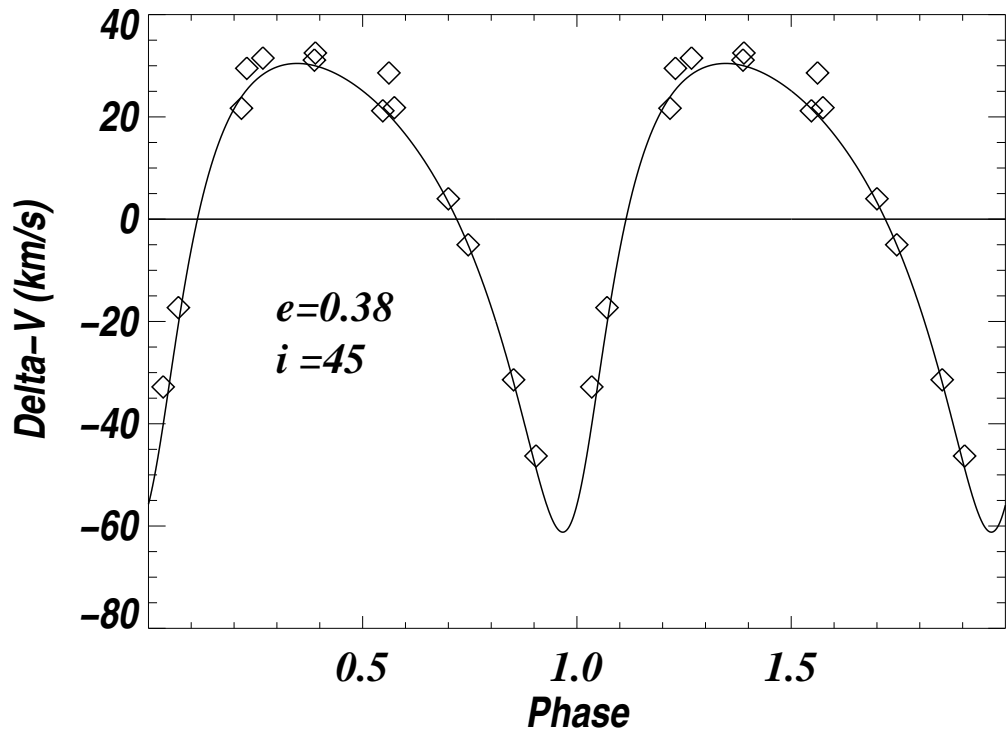,width=6in}}
\caption{One solution to the orbit of PPl 15. The orbital parameters are as given in Table~2, except as indicated on the Figure. This has a mean residual error in the solution for \delv~ of 2.8 \kms.}
\end{figure}

\begin{figure}{\label{orbit2}}
\centerline{\psfig{figure=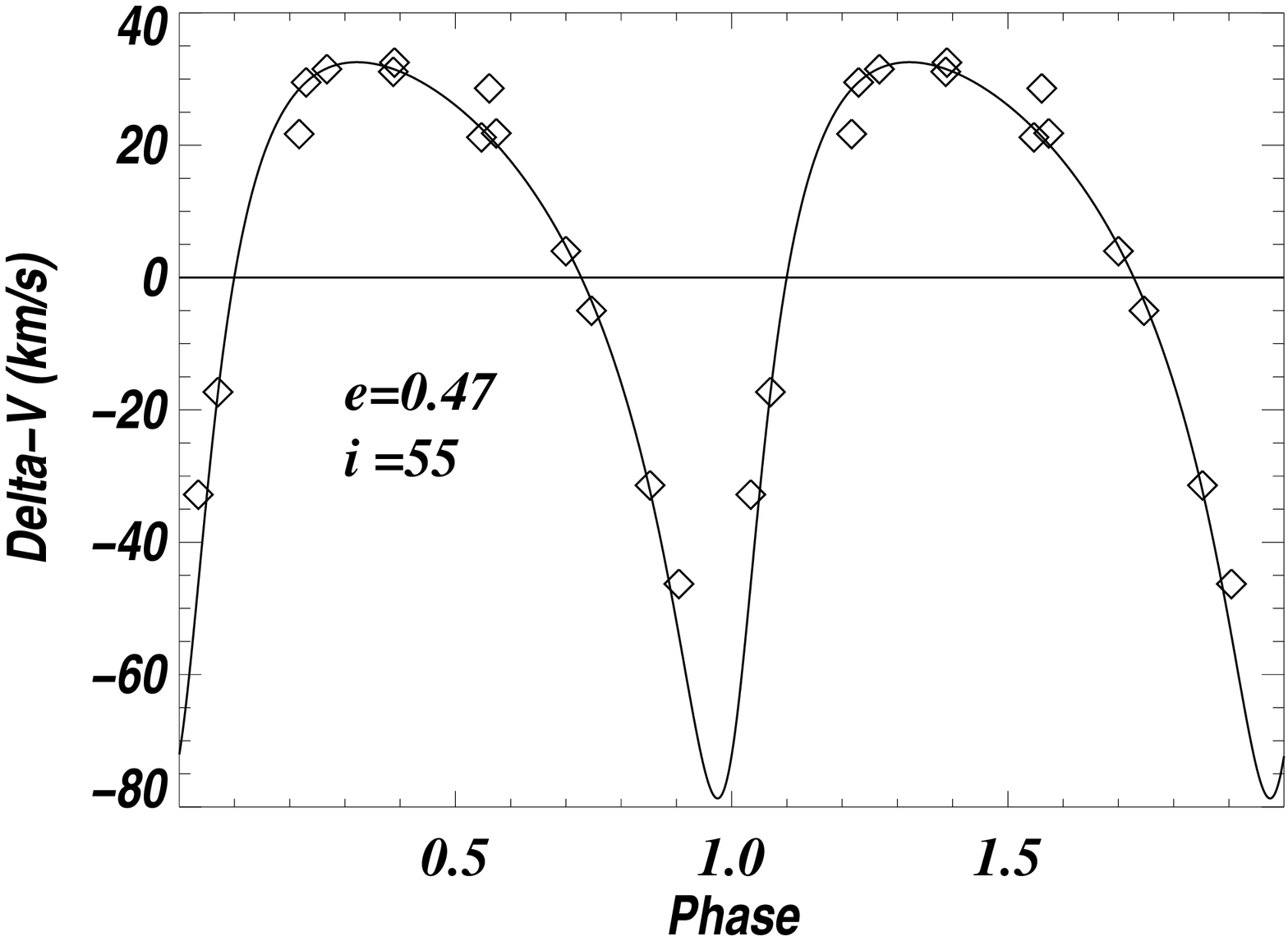,width=6in}}
\caption{Another good solution to the orbit of PPl 15. It differs primarily in the eccentricity and inclination, as indicated. This has a mean residual error in the solution for \delv~ of 3.1 \kms.}
\end{figure}

\subsection{The orbit of PPl~15}

To find the orbit we worked directly from the measured velocity differential 
between the two components for each night. It was obvious that the orbital 
period is about a week from the recurrence interval of the unsplit profiles 
\citep{basmar98}, and the fact that the velocities do not change rapidly within 
one night (which we tested for in Dec. 1997). We used the narrowest single 
profile to set the orbital epoch for \delv~ = 0  and then computed a set of 
orbits for a variety of periods in increments of 0.01 days. With initial guesses 
for the orbital parameters $e,i,\omega$ and an assumption of mass ratio 0.85 and 
total mass 0.13M$_\odot$~ (based on the system luminosity) we calculated the 
residual difference between the predicted and observed \delv~ to find at which 
periods it was minimized. Because of the nature of the time sampling of our data 
set, there are a series of minima between about 5.5-6.5 days which are not very 
different from each other. And of course the differences between them depend on 
the orbital parameters chosen. We attempted to automate a procedure for 
discovering the true best orbit, but found that manual exploration was able to 
find superior solutions. The fact is that although we can obtain very nice 
orbital fits with the current data set, they are apparently not sufficient for 
determining the unique best orbit.

The best period we could find is 5.825 days. We adopt this period, and a mass 
ratio of 0.85 (based on the ratio of line depths discussed above). The other 
orbital parameters were refined by manual searching, until the fit produced in 
Fig. \ref{orbit1} was found. With equal weight to all observations, the lowest 
residual is found for that solution (with the mean difference between the 
observed and predicted \delv~ of 2.8 \kms). Fig. \ref{orbit2} shows an alternate 
solution which appears as attractive (with residual of 3.1 \kms). They differ 
primarily in the orbital eccentricity and inclination. From tests like these, we 
produce the final orbital parameters listed in Table~2, where the errors reflect 
the uncertainty found in solutions with mean residuals that were not larger than 
7 \kms. Note that the values in Table~2 do not necessarily correspond to either 
of the shown solutions, but are averages with errors for each parameter. We do 
not claim to have explored all of parameter space, nor that these orbital 
parameters would hold up to extensive further observations. It is unlikely, 
however, that they would become different enough to affect any conclusions we 
draw below. The $M$ sin$i$ for the 2 components are 49-57\mj~ for component A 
and 42-48\mj~ for component B.

Our listed inclinations are based on the assumed masses from \S3.1 and the 
orbital period -- the observed velocities then imply an inclination. This is not 
a direct $dynamical$ demonstration of the substellar nature of PPl~15. The fact 
that we find very reasonable orbital solutions shows that the substellar masses 
deduced from luminosity are consistent with the dynamical appearance of the 
system. What is clear is that since the system appeared to be at the substellar 
luminosity boundary when thought of as a single object, breaking it into two 
components insures that each one is definitely a brown dwarf. 

\subsection{The lithium abundance in PPl~15}

We also reexamined the issue of lithium in PPl~15.  The announcement of its
detection at the AAS meeting in June 1995 (Science News 147, p.389) was the
first public declaration of discovery of a brown dwarf that did not
subsequently have to be recanted.  \citeauthor{bmg} expressed the desire for further confirmation of the line, and of the cluster membership of PPl~15. Of course they did not realize that the object was an SB2.  Proper motion confirmation of cluster membership has come from \citet{hambly99}. Our observations of Dec. 1997 were for the purpose of finding the orbit, not detecting lithium (which is at the blue end of our spectra and requires longer exposures). Nonetheless, we can add up all the spectra, recognizing that sometimes the lithium line is single and usually is split by varying velocities, to see if there is a depression in the spectrum as expected.  We find that there is, and the lithium detection is confirmed.

We can now understand better the results in \citeauthor{bmg}. On their first night the lines were not split, and the lithium feature is most obvious (they mention that 23 Nov 1994 provided the best data). In their Mar. 1995 data, not only was the moonlight contamination more severe, but the lines were quite split. In the case of very low signal-to-noise spectra, splitting the lines means essentially twice the noise for the overall lithium feature. This reduces the apparent equivalent width of the feature. The published lithium feature in \citeauthor{bmg} was also asymmetric. They thought that due merely to noise, but it turns out that modeling the line profile summed over the phases they added shows the asymmetry can actually produced by the SB2 nature of the object. This is further confirmation that their detection was robust and that both components have lithium, but we cannot reliably measure the lithium in each component. 

The accuracy of measurements is compromised by the necessity of adding spectra 
of various splittings (with the extra noise and continuum included). The 
spectrum is dominated by the hotter component, which should have an intrisically 
weaker lithium line for a given abundance. Finally, the models 
\citep{allhaus95} on which the depletion was based have since been revised, so 
the analysis should be done again. Thus, \citeauthor{bmg}'s conclusion that lithium is partially depleted in PPl~15 must be recanted; the measurements are not really good enough to say. Indeed, that would be surprising given that the individual components are now lower in mass and should not have depleted lithium. 

\subsection{Variability and flaring in \hal}

The \hal~ line was present in each of our echelle spectra. As noted in 
\citeauthor{bmg}, there is a nebular contribution to the line as well as stellar 
emission, but this is subtracted out along with the night sky lines. Thus we 
have an equivalent width and line profile for each observation, with the 
exception of 2-4 Dec 1997 during which observing conditions were sufficiently 
poor that although the \hal~ emission is seen, we cannot measure the underlying 
continuum and so cannot measure the true strength of the line. On other 
occasions the continuum is barely visible, so our measure of the line strength 
is rather inaccurate. The line varies between about 7-14\AA~ equivalent width on 
most nights; there is definitely some intrinsic variability in addition to the 
measurement errors (which we estimate at about 2\AA~ in typical cases). 

\begin{figure}{\label{hafig}}
\centerline{\psfig{figure=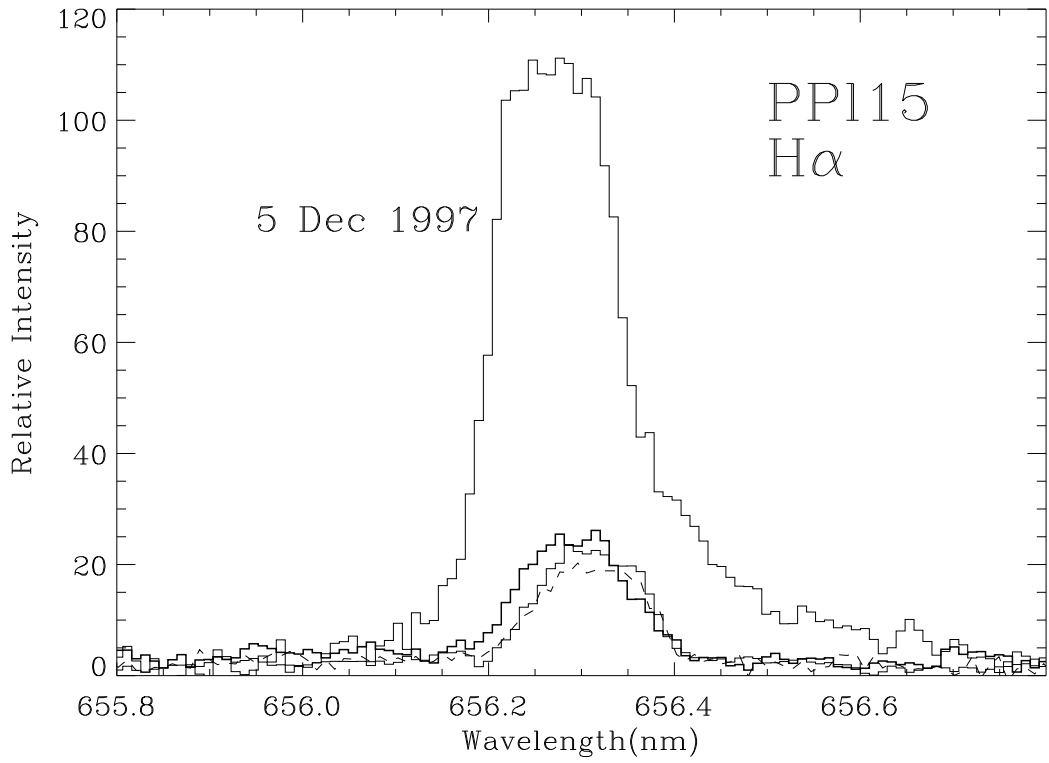,width=6in}}
\caption{The average \hal~ profiles of PPl~15 from each of the 3 epochs: Nov. 1994 (thick), Mar. 1995, Dec. 1997 (dashed, excluding the flare) are shown as the lower curves. The flare profile from 5 Dec  1997 is the much brighter line. }
\end{figure}

On 5 Dec 1997 there is a large flare in the system. The equivalent width of the 
line increases by a factor of 4-5 over its typical value, and the line develops 
a pronounced high-velocity red wing, extending to 200 \kms. This is shown in 
Fig. \ref{hafig}, along with the average non-flare profiles from our 3 observing 
epochs. The orbital phase at this time is 0.9, so the stars are approaching each 
other. In addition to \hal, we see HeI (667.8 nm) and CaII (850,866 nm) 
emission. These lines are symmetric and quite narrow, and have equivalent widths 
of 2.2, 1.1, and 1.2 \AA~ respectively. They appear to be associated with the 
primary only.

It is possible that the flare is associated with a magnetospheric interaction 
between the stars, as has been seen in RS CVn systems \citep[UX Ari;] []{simon80}. Indeed, in that case red wings were also seen, extending to 450 \kms~ in the MgII resonance lines. Their interpretation was in terms of a flow between the stars through interacting magnetic loops. The close T Tauri binary  \citep[DQ Tau;][]{basri97} also displays a tendency to ``flare'' as the stars approach, but in that case it has been interpreted as more likely due to an increase in the accretion activity on those young stars. That explanation should not apply here. Of course, it is possible that the orbital phase has nothing to do with the flare; with only one event we make no claim that there is such a connection. Low mass stars are certainly known to flare even when isolated. Our observation suggests that photometric monitoring of the PPL 15 system would be a fruitful project, although we do not know if the \hal~ flare was accompanied by a photometric change.

\section{Discussion}

\subsection{The eccentricity of PPl~15} 

The tidal circularization theory of \citet{zahn89} demonstrated that
close binaries with masses ranging from 1.25 to 0.5~M$_\odot$ and 
critical periods (P$_C$) less than about 8 days would have
circularized during their early pre-main sequence evolution.  Prior to PPl~15,
there were only two Pleiades binaries known with periods shorter than
critical, namely HII761 (P=3.31 days) and HII2407 (P=7.05 days).  Their orbits
are observed to be circular \citep{merm92}.  PPl~15 has a shorter
period than HII2407 yet it has a large eccentricity.  According to the models
of \citet{zahn89}, the value of P$_C$ depends primarily on the
pre-main sequence parameters (masses, radii and temperatures) of the binary
components, because the tidal interaction is maximized for larger radii and
larger convection zones (in this case the stars have remained fully convective). 
 The fact that PPl~15 is also relatively young
compared to binaries in the field is considered less relevant, since the
circularization occurs essentially in the pre-main sequence phase.

For equal mass binaries with initial $e$=0.3 and components of 1.0~M$_\odot$ and 
0.5~M$_\odot$, they found P$_C$ on the main sequence of 7.80 and 7.28 days, 
respectively.  Binaries with unequal mass components have slightly 
shorter critical periods.  The calculations agree well with the circular orbit  
of HII2407 for the estimated masses of its components (0.82~M$_\odot$ and 
0.27~M$_\odot$ \citep{merm92}. Extrapolating Zahn \& Bouchet's equation 14 and 
using the conditions for the `birthline' of deuterium burning provided by 
\citet{stahler88}, we infer that a binary with equal components of 0.1~M$_\odot$ 
has P$_C \sim$4.5 days.  The observed period of PPl~15 is longer than this 
value, and hence the binary is not expected to circularize. This is consistent 
with the eccentricity of our orbital solution. 

\subsection{Brown dwarf binaries} 

 The observed frequency of very short period binaries (P$<$10 days) is about
2\% for main sequence G-dwarfs and for G- and K-type Pleiades members
\citep{merm92}.  For X-ray selected stars in star forming regions the frequency 
of short period binaries increases by about a factor of 2
\citep{mathieu96} because of a selection bias (close binaries are brighter and 
more X-ray active than single stars).  In flux-limited CCD surveys like that of 
\citet{stauffer94}, there is also a slight bias to detect binaries
because they are brighter than single stars.  However, it is surprising to
find that the first one selected from that type of survey has a period of
only 5.8 days, because this is a rare type of binary.  The distribution of
binaries as a function of period has a broad peak at periods from 10$^4$ to
10$^5$ days \citep{duqmayor91}.  \citet{mart99} resolved one binary out of six
Pleiades BD candidates using HST imaging.  However, we have obtained
follow-up spectroscopy that does not support the membership of this binary in
the Pleiades.  

We have not found any binaries with separations $\ge$40~AU in a sample of 
22 Pleiades BD candidates observed with HST (these observations will be 
presented in another paper). If the distribution of binary frequencies among 
Pleiades BDs was similar to those of young stars and G dwarfs, we should have 
found 4.5 binaries.  On the other hand, the existence of PPl~15 indicates that 
BD binaries may not be rare (since it was the first Pleiades BD examined for 
binarity).  Four more pieces of evidence support this conclusion.  First, 
\citet{martbran99} have found one field BD binary out of two targets examined.  
The separation of this binary is about 5~AU.  Second, the 2MASS and DENIS 
surveys have not reported any visual binaries among their L dwarfs 
\citep[though there are preliminary reports of other close binaries similar to that in ]{martbran99}. Third, color-magnitude diagrams of the Pleiades VLM stars 
and BDs show a large spread that has been interpreted as being due mainly to 
unresolved binaries \citep{steele93,osorio97}.  Fourth, the presence of 
an unresolved substellar secondary has been inferred from infrared spectroscopy 
of the VLM Pleiades star HHJ54 \citep{steele95}. There is no evidence for other 
objects in the PPl~15 system of similar mass to those in the close binary either 
from photometry \citep{osorio97} or HST imaging. 

If BD binaries are not rare, the fact that they have not been found at 
separations larger than about 40 AU in the Pleiades and the field suggests that 
the distribution of binary separations may be different than that of young stars 
and G dwarfs.  One possible difference is that wide substellar binaries are less 
tightly bound than their stellar counterparts, and are therefore easier to 
disrupt by gravitational interactions.  This could account for the lack of BD 
binaries with separations $\ge$40 AU, but it would not increase the frequency of 
short period binaries.  The formation of PPl~15 should still be an unlikely 
event if the intrinsic distribution of separations is at all similar to that for 
stars (even if the distribution has a smaller overall scale).  In order to 
explain the fact that PPl~15 has a very short period along with the lack of 
Pleiades BDs with separations $\ge$40~AU, we propose that the formation of 
substellar binaries tends to produce smaller separations than the formation of 
stellar binaries. To the best of our knowledge, this effect has not been 
explicitly predicted by any binary formation model.

Reviews of theories of binary formation (stellar and substellar) can be found in 
\citet{boden98} and \citet{bodenpp4}. There is currently no consensus theory for 
the formation of very close binaries, either stellar or substellar. If they form 
in situ, it could depend on the inability of a collapsing core to resolve its 
angular momentum problem just at the last moments before it would form a single 
core. PPl~15 has a mass ratio close to unity, suggesting fragmentation of a bar 
(not too dissimilar from the now discredited fission hypotheses for binary 
formation). Formation of substellar objects in bars and filaments is seen in 
some of the calculations \citep{bodenpp4}. Production of a second object in the 
circumstellar disk around an already formed primary is unlikely to apply to this 
case; the secondary is too close and too similar in mass to the primary.

Alternatively, the system we see now may not directly reflect its inital 
formation. Models of direct fragmentation of a collapsing cloud can produce 
nearly equal mass components with large eccentricities. However, the typical 
separations are in the range 10--100 AU \citep{boss86}. If the PPl~15 system 
formed in a small association of protostars and was subsequently ejected from 
the group, the ejection velocity would have to be low because the system remains 
bound to the open cluster.  One possibility is that it formed as a triple 
substellar system, and the tight pair was produced at the expense of the 
ejection of the third component.  

Models of multiple fragmentation yield complicated structures.  \citet{burk96} 
have made simulations where they find an inner BD binary with a separation of 
about 20~AU, which induces formation of an outer BD binary with separation of 
about 100~AU.  Such a configuration is unstable, and orbital evolution of the 
fragments occurs, but it has not been followed until more than 10\% of the cloud 
mass was accreted. It is interesting to note that orbital evolution has gained 
great favor in explaining the orbits of the extrasolar giant plants found by 
radial velocity techniques \citep{marcy98}.

The short period of PPl~15 is both observationally and theoretically surprising. 
No systematic search for other spectroscopic BD binaries has been made yet, 
although we and others are starting them. Explaining the formation of PPl~15 
provides a new challenge for models of (substellar) binary formation. In any 
event, the discovery of the multiplicity of PPl~15 initiated the study of binary 
systems composed solely of substellar objects.
 
\acknowledgments

{\it Acknowledgments}: 
This research is based on data collected at the  W.~M. Keck Observatory, which 
is operated jointly by the University of  California and California Institute of 
Technology. We acknowledge the support of NSF through grant AST96-18439. GB was 
a Miller Research Professor during this project. EM acknowledges support from 
the Fullbright-DGES program  of the Spanish Ministry of Education. We thank 
again G. Marcy, P. Butler, J. Stauffer, and R. O'Dell for their cooperation 
during the Dec. 1997 observing run.


\begin{deluxetable}{llccc}
\footnotesize
\tablecolumns{5}
\tablewidth{0pc}
\tablecaption{Observations of PPl 15}
\tablehead{
\colhead{JD(24+)}  & 
\colhead{UT date}  &
\colhead{$\Delta$v}  & 
\colhead{EW (H$\alpha$)}  &
\colhead{Orbital Phase}  \\  }
\startdata
49679.92 & 23 Nov 1994 & 4.0   & 15 & 0.70 \\
49680.81 & 24 Nov 1994 & -33.5 & 7  & 0.85 \\
49681.87 & 25 Nov 1994 & -35.8 & 14 & 0.04 \\
49788.79 & 12 Mar 1995 & 33.1  & 8  & 0.39 \\
49789.79 & 13 Mar 1995 & 28.5  & 13 & 0.56 \\
50783.86 & 01 Dec 1997 & 21.8  & 12 & 0.22 \\
50784.85 & 02 Dec 1997 & 31.2  & \nodata  & 0.39 \\
50785.783& 03 Dec 1997 & 21.2  & \nodata  & 0.55 \\
50785.938& 03 Dec 1997 & 21.8  & \nodata  & 0.57 \\
50786.94 & 04 Dec 1997 & -5.0  & \nodata  & 0.75 \\
50787.87 & 05 Dec 1997 & -46.5 & 48 & 0.91 \\
50788.83 & 06 Dec 1997 & -17.3 & 15 & 0.07 \\
50789.757& 07 Dec 1997 & 29.5  & 11 & 0.23 \\
50789.977& 07 Dec 1997 & 31.5  & 11 & 0.27 \\
\enddata
\tablenotetext{a}{$\Delta$v is the velocity separation of the two components in 
km s$^{-1}$ (a negative sign indicates the components are reversed); the 
emission equivalent widths of H$\alpha$ are given in \AA~ (Dec. 2-4 were 
insufficiently well exposed to measure).}
\end{deluxetable}

\begin{deluxetable}{ll}
\footnotesize
\tablecolumns{2}
\tablewidth{280pt}
\tablecaption{Orbital Parameters for PPl15}
\tablehead{
\colhead{~~~~~~~~~~~~}  & 
\colhead{~~~~~~~~~~~~~~~~~~~~~~~~}  \\ }
\startdata
$P$ (days) & $5.825 \pm 0.3$ \\
$a$ (AU) & 0.03 \\
$e$ & $0.42\pm 0.05$ \\
$i$  & $50^\circ \pm 5^\circ$ \tablenotemark{a} \\
$\omega$  & $62^\circ$ \\
Epoch	& $2450782.59 \pm 0.01$ \\
M$_2$/M$_1$  & $0.85 \pm 0.05$ \tablenotemark{b} \\
M$_1$, M$_2$ (M$_{Jup}$) & 70, 60 ($\pm$3) \tablenotemark{c} \\
\enddata
\tablenotetext{a}{ The inclination and semimajor axis follow from the assumed 
masses, which are luminosity based.}
\tablenotetext{b}{ The mass ratio is found from spectral line depth ratios.}
\tablenotetext{c}{The sum of the masses is estimated from the total system 
luminosity using theoretical models.}  
\end{deluxetable}

\end{document}